# Estimating regional cerebral blood flow using resting-state functional MRI via machine learning

Ganesh B Chand[1,3,*], Mohamad Habes[1,2,3], Sudipto Dolui[1,2], John A Detre[1,2], David A Wolk[2], Christos Davatzikos[1,3,*]

[1]Department of Radiology, Perelman School of Medicine, University of Pennsylvania, Philadelphia, Pennsylvania, USA
[2]Department of Neurology, Perelman School of Medicine, University of Pennsylvania, Philadelphia, Pennsylvania, USA
[3]Center for Biomedical Image Computing and Analytics, Perelman School of Medicine, University of Pennsylvania, Philadelphia, Pennsylvania, USA

## Abstract

Perfusion MRI is an important modality in many brain imaging protocols, since it probes cerebrovascular changes in aging and many diseases; however, it may not be always available. Here we introduce a method that seeks to estimate regional perfusion properties using spectral information of resting-state functional MRI (rsfMRI) via machine learning. We used pairs of rsfMRI and arterial spin labeling (ASL) images from the same elderly individuals with normal cognition (NC; n = 45) and mild cognitive impairment (MCI; n = 26), and built support vector machine models aiming to estimate regional cerebral blood flow (CBF) from the rsfMRI signal alone. This method demonstrated higher associations between the estimated CBF and actual CBF (ASL-CBF) at the total lobar gray matter (r = 0.40; FDR-p = 1.9e-03), parietal lobe (r = 0.46, FDR-p = 8e-04), and occipital lobe (r = 0.35; FDR-p = 0.01) using rsfMRI signals of frequencies [0.01-0.15] Hertz compared to frequencies [0.01-0.10] Hertz and [0.01-0.20] Hertz, respectively. We further observed significant associations between the estimated CBF and actual CBF in 24 regions of interest (p < 0.05), with the highest association observed in the superior parietal lobule (r = 0.50, FDR-p = 0.002). Moreover, the estimated CBF at superior parietal lobule showed significant correlation with the mini-mental state exam (MMSE) score (r = 0.27; FDR-p = 0.04) and decreased in MCI with lower MMSE score compared to NC group (FDR-p = 0.04). Overall, these results suggest that the proposed framework can obtain estimates of regional perfusion from rsfMRI, which can serve as surrogate perfusion measures in the absence of ASL.

**\*Correspondence:** Ganesh B Chand, PhD (ganesh.chand@uphs.upenn.edu) or, Christos Davatzikos, PhD (christos.davatzikos@uphs.upenn.edu)
University of Pennsylvania, 3700 Hamilton Walk, Richards Building, 7th Floor, Philadelphia, PA 19104. Tel: +1 215-746-4060

## 1. Introduction

Arterial spin labeled (ASL) perfusion magnetic resonance imaging (MRI) can noninvasively quantify regional cerebral blood flow (CBF), which is an important physiological parameter of brain function and a versatile biomarker in aging and brain disorders, including cerebrovascular disease, Alzheimer's disease, and neuropsychiatric disorders (Bruandet et al., 2009; Detre et al., 2012; Gupta et al., 2012; Haller et al., 2016; Love and Miners, 2016; MacDonald and Frayne, 2015). Although ASL MRI can be used to measure task activation, most ASL studies are carried out in the "resting" state and are intended to address state- or trait-like aspects of brain function. Task-independent blood oxygenation level dependent (BOLD) functional MRI (fMRI), often termed resting-state fMRI (rsfMRI) is another common strategy for evaluating state- and trait-like properties of brain function.

ASL uses magnetically labeled arterial blood water as an endogenous tracer (Detre et al., 1992) to quantify CBF in physiological units of ml/100g/min. ASL MRI has been validated against $^{15}$O-PET (Fan et al., 2016; Feng et al., 2004; Kilroy et al., 2014; Ye et al., 2000) and has demonstrated sensitivity to regional CBF alterations associated with Alzheimer's neurodegeneration (Alsop et al., 2010; Dai et al., 2009; Johnson et al., 2005; Wang et al., 2013; Xie et al., 2016). Further, based on the tight coupling between regional CBF and regional brain metabolism, regional hypoperfusion by ASL and regional hypometabolism by $^{18}$F-fluorodeoxyglucose positron emission tomography (FDG-PET) display similar patterns (Chen et al., 2011; Verfaillie et al., 2015) suggesting the potential of ASL to provide noninvasive and less costly surrogate marker than FDG-PET in aging, MCI and/or Alzheimer's disease (Wolk and Detre, 2012). However, robust acquisition procedures and consensus on best practices in ASL MRI have only recently begun to be available (Alsop et al., 2015) and ASL remains technically more challenging than BOLD fMRI.

RsfMRI has emerged as a commonly acquired MRI protocol primarily used in studies of brain functional connectivity (Biswal et al., 1995; Biswal et al., 2010; Chand et al., 2018; Fox et al., 2014; Greicius and Kimmel, 2012; Raichle, 2015). It indirectly captures the regional neurovascular alterations associated with neuronal activations based on associated changes in deoxyhemoglobin reflecting changes in underlying regional CBF, blood volume, and oxygen metabolism. Low frequency oscillations recorded in rsfMRI are typically investigated between 0.01 and 0.10 Hertz (Fox et al., 2005; Fransson, 2005). Emerging studies have begun to reveal that perfusion properties may be extracted from the rsfMRI time series (Lv et al., 2013; Tong and Frederick, 2014; Tong et al., 2017; Yan et al., 2018). Because dynamic changes in BOLD contrast are primarily due to changes in CBF (Detre et al., 2012; Li et al., 2012; Raichle, 1998), rsfMRI has also been explored as a potential alternative means of estimating cerebral perfusion (Tong et al., 2017; Yan et al., 2018). Those investigations suggest that appropriate filtering and modeling of the rsfMRI signals allow inferences to be made about regional perfusion, which would be of particular value when ASL MRI data is not available. Herein we propose a new method to estimate perfusion properties from the rsfMRI using a machine learning model.

In the current study, we *hypothesized* that CBF metrics can be estimated from rsfMRI signals at an array of different frequencies, using machine learning models constructed from individuals scanned with both protocols and therefore serving as a training set. We further *hypothesized* that the predicted regional CBF metrics correlate with global cognitive performance.

## 2. Materials and Methods

### 2.1. Study sample

Subjects were recruited from the Penn Memory Center/Penn Alzheimer's Disease Core Center (ADCC). Institutional Review Board of the University of Pennsylvania reviewed and approved the study protocol. All participants receive annual assessments as part of their participation in the ADCC, which includes psychometric testing prescribed by the Uniform Data Set (Dolui et al., 2017).There were 71 older adults who completed both rsfMRI and ASL scans. In the total sample, 45 individuals were cognitively normal and 26 carried a diagnosis of mild cognitive impairment (MCI). The clinical diagnosis of MCI was determined based on a consensus discussion by neurologists, psychiatrists, geriatricians, and neuropsychologists following the criteria outlined by Petersen and others (Petersen, 2004; Petersen et al., 2009; Winblad et al., 2004). Study exclusion criteria were a history of clinical stroke, significant traumatic brain injury, alcohol or drug abuse, or any medical or psychiatric condition that thought to impact cognition. Sample demographics are presented in **Table 1**.

**Table 1:** Sample demographics

| Variable | Normal controls (NC) (n = 45) | MCI (n = 26) | P value |
|---|---|---|---|
| Age, mean years (SD) | 70.56 (7.21) | 70.50 (6.11) | 0.23 |
| Sex, female (%) | 27 (60.00%) | 11 (42.31%) | 0.15 |
| Education, mean years (SD) | 16.18 (2.78) | 16.38 (2.77) | 0.76 |
| MMSE, min-max (SD) | 26-30 (0.87) | 22-30 (2.68) | 3.09e-09 |

### 2.2. Image acquisition

MR images were obtained using Siemens 3T Prisma scanner at the University of Pennsylvania, Philadelphia. Anatomical 3D T1 images were acquired in sagittal using the repetition time (TR) = 2400 ms, echo time (TE) = 2.24 ms, inversion time (TI) = 1060 ms, flip angle = 8 degree, matrix = 320 x 320, slices per slab = 208, and slice thickness = 0.80 mm. The rsfMRI images were acquired using the TR = 720 ms, TE = 37 ms, flip angle = 52 degree, field of view (FOV) = 208 x 208 mm$^2$, matrix = 104 x 104, number of slices = 72, slice thickness = 2 mm, and bandwidth = 2290 Hertz/pixel. ASL images were acquired with unbalanced pseudo-continuous labeling with a labeling time of 1.8 s, post labeling delay of 1.8 s and performed at an optimal location perpendicular to straight segments of the internal carotid and vertebral arteries as determined by time-of-flight angiography. Image readout consisted of 1D-accelerated 4-shot 3D stack-of-spirals acquisition with 90% background suppression using TR = 4250 ms, TE = 9.8 ms, TI = 150 ms, flip angle = 90 degree, FOV = 240 x 240 mm$^2$, matrix = 96 x 96, slices per slab = 52, and voxel size

= 2.5 x 2.5 x 2.5 mm$^3$. Ten label/control pairs were acquired for signal averaging. Two volumes of M$_0$ images acquired without background suppression were averaged and used to normalize the control-label difference for CBF quantification.

## 2.3. RsfMRI preprocessing, time series extraction, and spectral features

We preprocessed rsfMRI using the SPM12 (Wellcome Trust Centre for Neuroimaging, London, United Kingdom; www.fil.ion.ucl.ac.uk/spm/software/spm12). Briefly, preprocessing steps consisted of slice time correction, motion correction, coregistration to an individual anatomical T1 image, normalization to Montreal Neurological Institute (MNI) template space, and spatial smoothing of normalized images using 6 mm isotropic Gaussian kernel. Total 117 anatomical gray matter regions of interest (ROIs) defined by a multi-atlas segmentation method (MUSE) (Doshi et al., 2013; Doshi et al., 2016) were considered. RsfMRI time series were extracted from MUSE ROIs and Fourier coefficients were used to compute the spectral features at those ROIs.

## 2.4. ASL preprocessing and CBF extraction

The T1 images were probabilistically segmented into gray matter (GM), white matter (WM) and cerebrospinal fluid (CSF) using the SPM12 and they were subsequently used to compute a brain mask containing GM, WM, and ventricular CSF (Dolui et al., 2017). Each subject's T1-weighted images were used to estimate a subject-specific non-linear warping to the MNI152 template using FMRIB's non-linear image registration tool (FNIRT) (Andersson et al., 2010). ASL processing steps consisted of first realigning the raw EPI images using the method by Wang (Wang, 2012). The mean EPI image, obtained by temporally averaging the realigned EPI time series, was coregistered to the structural image using FSL boundary-based registration (BBR). Pairwise subtraction of the label-control images was performed and the difference was converted to absolute CBF time series measurements using a single compartment model (Alsop et al., 2015) with parameters: brain/blood partition coefficient = 0.9, $T_{1, blood}$ = 1664 ms. Subject's CBF maps were normalized to the MNI152 space using a subject-specific ASL-to-MNI152 warp obtained by concatenating the ASL-to-T1 BBR coregistration parameters with T1-to-MNI152 warp. The CBF values were finally extracted using MUSE-defined anatomical ROIs and lobes.

## 2.5. CBF prediction

We binned the spectral information—the square of Fourier coefficients referred to as spectral power—of rsfMRI into 8 bins, which covered 0.01-0.20 Hertz with 0.0268 Hertz separation between consecutive bins. To examine the prediction with varying frequency range, the rsfMRI signals were further taken into [0.01-0.10] Hertz, [0.01-0.15] Hertz, and [0.01-0.20] Hertz. We selected signals [0.01-0.10] Hertz based on a typical frequency range included in rsfMRI studies (Fox et al., 2005; Fransson, 2005). The signals [0.01-0.15] Hertz were selected as the recent studies have begun to show that rsfMRI signals with this frequency range might have higher perfusion information (Tong et al., 2017). The frequency range [0.01-0.20] was chosen to compare with the former two ranges. However, frequency ranges above 0.2 Hertz were not included as the signals might pick up the physiological noises such as cardiac cycles (>0.5 Hertz) and respiratory cycles (~0.2-0.3 Hertz) (Dagli et al., 1999; Limbrick-Oldfield et al., 2012; Raj et al., 2001).

These rsfMRI signals were used as features in the kernel-based support vector machines (Cortes and Vapnik, 1995) to predict the CBF. The support vector machines regression was performed using libSVM package (Chang and Lin, 2011; Drucker et al., 1997) based on 10 folds cross-validation. Briefly, the data was divided into 10-subsets and the method was repeated 10-times. Each time, 9-subsets were used as training set and the remaining 1-subset as a testing set. Gaussian kernel function was used in the model and model parameters were optimized within each training set using 10 folds cross-validation grid search.

### 2.6. Data and/or code availability statement

The codes used in the present study will be made available in the public domain as a software package.

### 2.7. Statistical analysis

The cognitively normal and MCI patients were compared using two-sample t-test regarding their age, education, and mini-mental state exam (MMSE). Sex was compared between two groups using a chi-square test. Age and sex were adjusted in CBF using a linear model. Pearson's correlation analyses were carried out to examine the associations between the CBF measured using ASL data and that predicted from rsfMRI data. The relative CBFs at superior parietal lobule and precuneus (i. e., normalized by lobar CBF) were compared between controls, MCI with higher cognitive score (MMSE ≥ 28), and MCI with lower cognitive score (MMSE < 28) using two-sample t-test. Pearson's correlation analyses were performed to examine the association between relative CBF and global cognitive performance assessed by MMSE. A MATLAB software version R2018a (https://www.mathworks.com) was used for analyzing the data. The p-values were corrected for multiple comparisons using the false discovery rate (FDR).

## 3. Results

### 3.1. Lobar CBF prediction

The CBF predictions were carried out at the lobar gray matter by using rsfMRI signals with frequencies [0.01-0.10] Hertz, [0.01-0.15] Hertz, and [0.01-0.20] Hertz. Association between the actual CBF and predicted CBF at the total lobar gray matter was higher using frequencies [0.01-0.15] Hertz (r = 0.40; FDR-p = 1.9e-03) than using frequencies [0.01-0.10] Hertz (r = 0.20; FDR-p = 0.09) and [0.01-0.20] Hertz (r = 0.26, FDR-p = 0.04), respectively, as shown in **Figure 1**. The correlation coefficients between predicted CBF and actual CBF varied as a function of frequency (Supplementary materials: **Figure S1**). In each subject, the absolute difference between actual CBF and predicted CBF showed fluctuating behavior with frequency (Supplementary materials: **Figure S2**). The mean (standard deviation) of absolute difference over all subjects versus frequency value is shown in **Figure S3** (Supplementary materials).

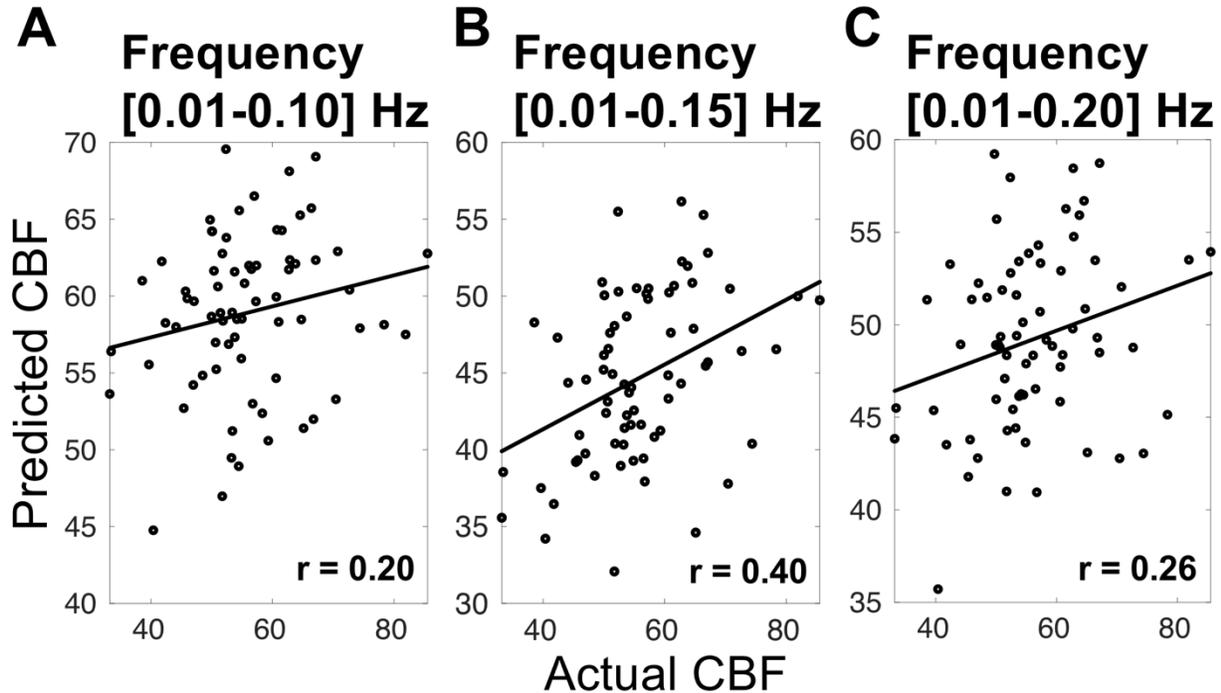

**Figure 1:** Associations between the actual CBF and predicted CBF at total lobar gray matter with varying frequency information of rsfMRI: **A)** frequency [0.01-0.10] Hertz (Hz) (r = 0.20, FDR-p = 0.09), **B)** frequency [0.01-0.15] Hz (r = 0.40, FDR-p = 1.9e-03), and **C)** frequency [0.01-0.20] Hz (r = 0.26, FDR-p = 0.04). Note that the association is higher in [0.01-0.15] Hz compared to that of other frequency ranges.

Prediction analyses were also performed at the frontal, temporal, parietal, and occipital lobes using rsfMRI with frequencies [0.01-0.10] Hertz, [0.01-0.15] Hertz, and [0.01-0.20] Hertz. **Table 2** lists the correlations between the predicted and actual CBF values for each lobe and each frequency range. The highest and most significant correlations were observed in the parietal lobe with frequencies [0.01-0.15] Hertz (r = 0.46, FDR-p = 8e-04) (**Figure 2**). The correlations between actual CBF and predicted CBF were lower at the occipital lobe compared to that of the parietal lobe. Moreover, associations between actual CBF and predicted CBF were not statistically significant at the frontal lobe and temporal lobe (FDR-p > 0.05) as presented in **Table 2**.

**Table 2:** Associations between the actual CBF and predicted CBF at individual lobar level using rsfMRI signals with frequencies [0.01-0.10] Hertz (Hz), [0.01-0.15] Hz, and [0.01-0.20] Hz (p values are FDR-corrected).

| Lobe | Frequency [0.01-0.10] Hz | Frequency [0.01-0.15] Hz | Frequency [0.01-0.20] Hz |
|---|---|---|---|
| Frontal | r = 0.19; p = 0.16 | r = 0.09; p = 0.41 | r = 0.14; p = 0.26 |
| Temporal | r = 0.19; p = 0.16 | r = 0.16; p = 0.24 | r = 0.14; p = 0.26 |
| Parietal | r = 0.25; p = 0.09 | r = 0.46; p = 8e-04 | r = 0.36; p = 0.01 |
| Occipital | r = 0.25; p = 0.09 | r = 0.35; p = 0.01 | r = 0.21; p = 0.16 |

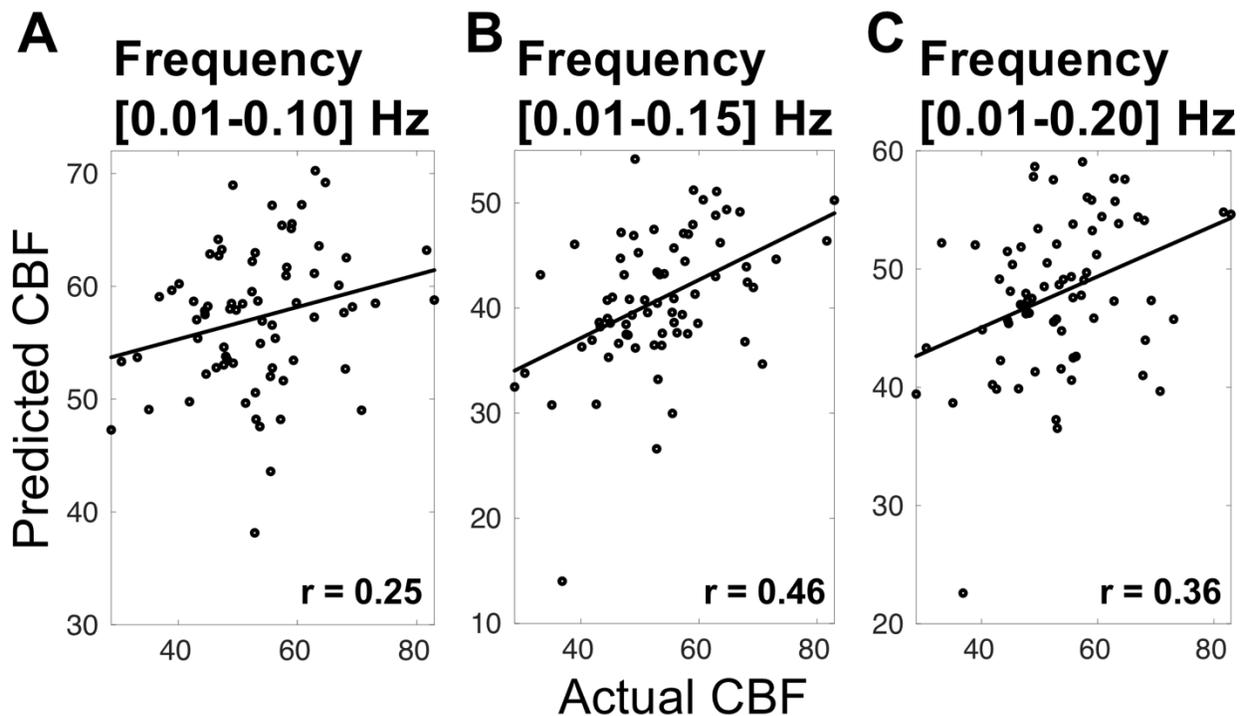

**Figure 2:** Associations between the actual CBF and predicted CBF at the parietal lobe with varying frequency information of rsfMRI: **A)** frequency [0.01-0.10] Hertz (Hz) (r = 0.25, FDR-p = 0.09), **B)** frequency [0.01-0.15] Hz (r = 0.46, FDR-p = 8e-04), and **C)** frequency [0.01-0.20] Hz (r = 0.25, FDR-p = 0.01). Note that the association is higher in [0.01-0.15] Hz compared to that of other frequency ranges.

### 3.2. ROI CBF prediction

We further carried out the prediction analyses at 117 MUSE ROIs. At each ROI level, the association between actual CBF and predicted CBF was highest using rsfMRI with frequencies [0.01-0.15] Hertz. Out of those ROIs, the correlations were statistically significant (uncorrected for multiple comparisons; p < 0.05) at 24 ROIs, which were predominantly the cortical ROIs (Supplementary materials: **Table S1**). After correcting for multiple comparisons, there were 7 statistically significant ROIs (FDR-p < 0.05) (Supplementary materials: **Table S1**). Predicted CBF at the subcortical ROIs, including putamen, hippocampus etc., was not statistically significant with actual CBF. Among 117 ROIs, the association was highest at the superior parietal lobule. **Figure 3** shows the association between actual CBF and predicted CBF at the superior parietal lobule using rsfMRI signals with frequencies [0.01-0.10] Hertz (r = 0.33, FDR-p = 0.09), [0.01-0.15] Hertz (r = 0.50, FDR-p = 0.002), and [0.01-0.20] Hertz (r = 0.34, FDR-p = 0.08) demonstrating better estimates in [0.01-0.15] Hertz.

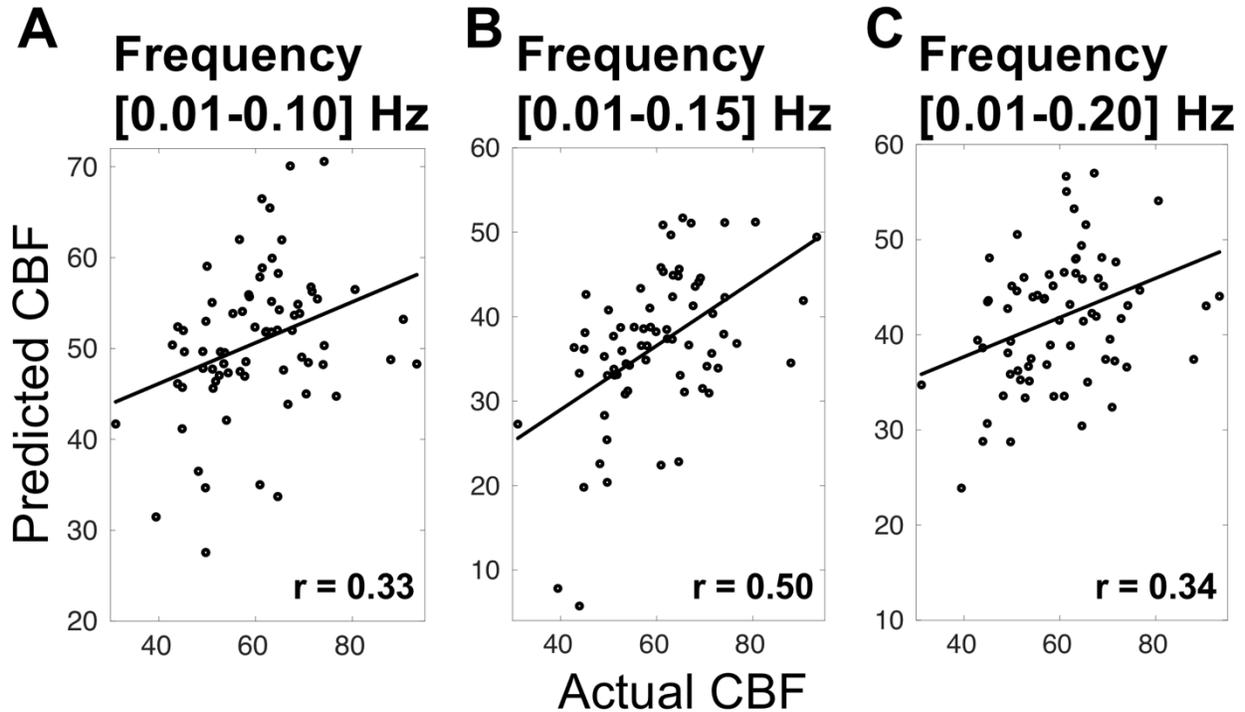

**Figure 3:** Associations between the actual CBF and predicted CBF at superior parietal lobule with varying frequency information of rsfMRI: **A)** frequency [0.01-0.10] Hertz (Hz) (r = 0.33, FDR-p = 0.09), **B)** frequency [0.01-0.15] Hz (r = 0.50, FDR-p = 0.002), and **C)** frequency [0.01-0.20] Hz (r = 0.34, FDR-p = 0.08). Note that the association is higher in [0.01-0.15] Hz as compared to that of other frequency ranges.

### 3.3 Predicted CBF comparison between controls and MCI

We further examined the predicted CBF at superior parietal lobule (i. e., normalized by predicted CBF of lobar) between controls, MCI with higher cognitive performance (MMSE ≥ 28), and MCI with lower cognitive performance (MMSE < 28). Similar to the actual CBF, the predicted CBF significantly decreased in MCI with lower cognitive performance (FDR-p < 0.05) compared to the controls (**Figure 4**). At precuneus, the similar trends were present between controls and MCI, but the p-values were not statistically significant.

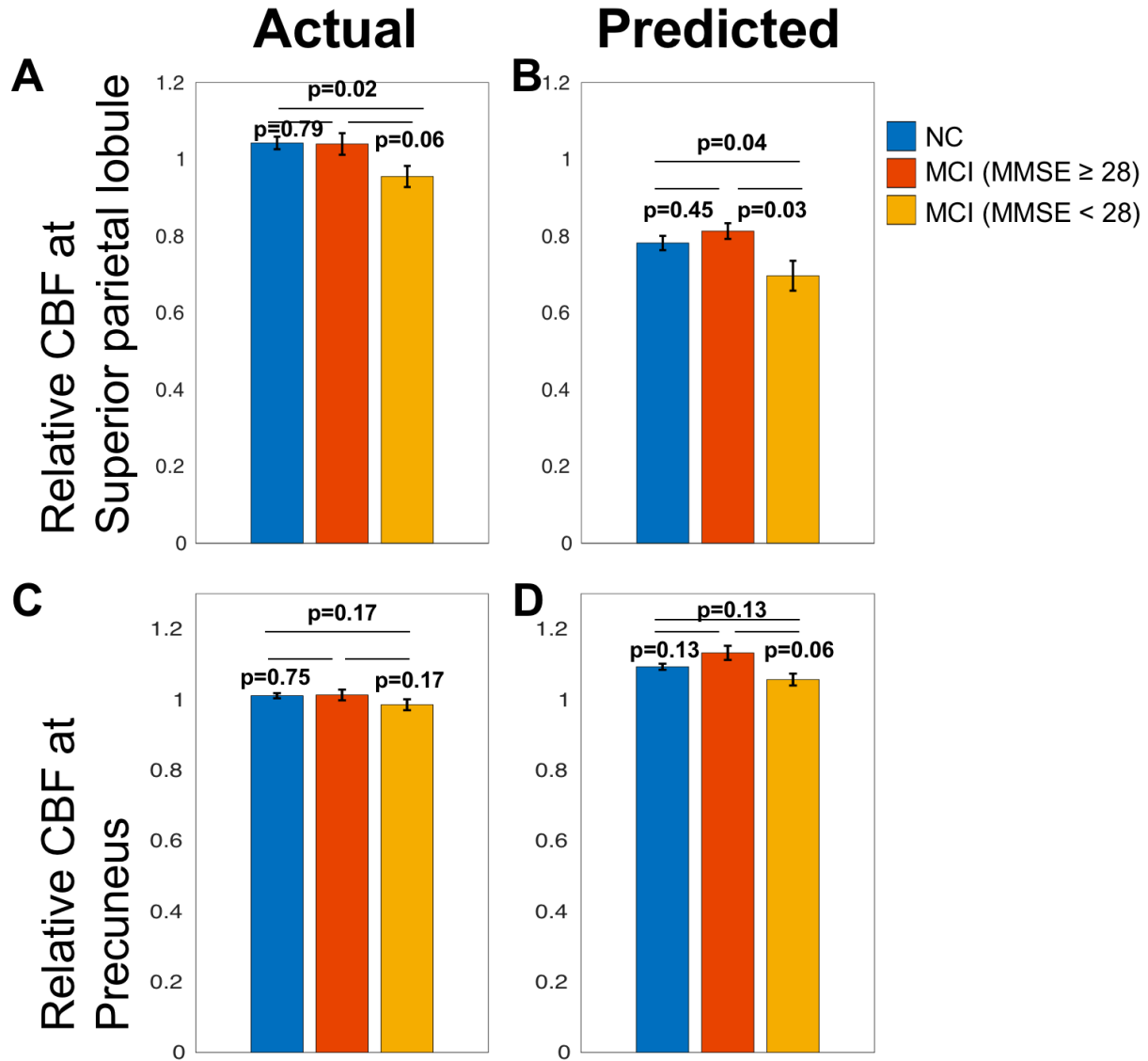

**Figure 4:** Comparing relative CBF between normal controls (NC), MCI with higher cognitive performance (MMSE ≥ 28), and MCI with lower cognitive performance (MMSE < 28): **A**) Actual and **B**) Predicted relative CBF at superior parietal lobule demonstrating a significant difference between NC and MCI with lower cognitive performance (MMSE < 28); **C**) Actual and **D**) Predicted relative CBF at precuneus demonstrating a trend of difference between NC and MCI with lower cognitive performance (MMSE < 28) (FDR-corrected p values within each region).

### 3.4. Associations between predicted CBF and cognition

We tested an association between predicted CBF and global cognitive performance assessed by MMSE (**Figure 5**). Similar to the actual CBF, the predicted CBF at superior parietal lobule correlated positively with MMSE (FDR-p < 0.05).

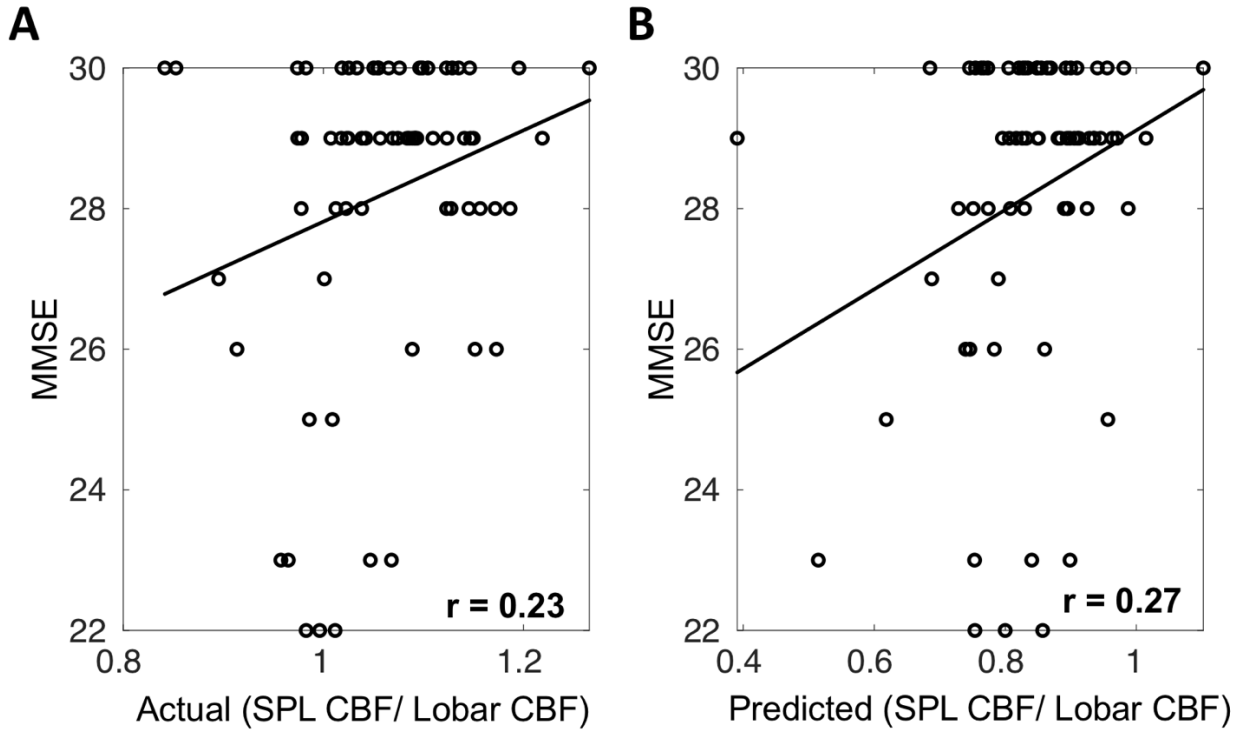

**Figure 5:** Associations between relative CBF at superior parietal lobule (SPL) and MMSE score: **A**) Correlation between actual relative CBF and MMSE (r = 0.23; FDR-p = 0.05), and **B**) Correlation between predicted relative CBF and MMSE (r = 0.27; FDR-p = 0.04).

## 4. Discussion

In the present study, we proposed a novel multi-folded framework that demonstrated that regional CBF metrics can be estimated using the spectral information of rsfMRI. In this approach, we acquired rsfMRI and ASL from same individuals, trained support vector machine regression models on those data using standard cross-validation techniques, predicted regional CBF measures, and finally illustrated an association of the predicted CBF with original (actual) CBF from the ASL data itself. Our analyses revealed that the predicted CBF at the lobar level and/or individual cortical ROI level associates higher with the respective actual CBF using rsfMRI signals of frequencies [0.01-0.15] Hertz. CBF estimates were better in the parietal lobe or region(s) than in other brain regions. Estimated CBF at the superior parietal lobule decreased in MCI group with lower MMSE compared to control group and associated positively with MMSE.

### 4.1. Better prediction using rsfMRI signals with frequencies [0.01-0.15] Hertz

The predicted CBF metrics varied as a function of frequency of rsfMRI signals. These results were in line with the reports that rsfMRI signal amplitude shows fluctuating behavior with the frequency (Buzsaki, 2004; Zuo et al., 2010). Our results provided strong evidence that the rsfMRI signals with frequencies [0.01-0.15] Hertz had higher associations between the predicted CBF and actual CBF at global lobar, individual lobe, and individual ROI compared to the signals of frequencies [0.01-0.10] Hertz and [0.01-0.20] Hertz, respectively. Perfusion estimations from rsfMRI were broadly in line with recent reports that rsfMRI signals may carry perfusion properties (Lv et al., 2013; Tong and Frederick, 2014; Yan et al., 2018). It has also been shown that the rsfMRI signals with [0.01-0.15] Hertz to a larger extent resembled perfusion measurement with dynamic susceptibility contrast compared to the signals with [0.01-0.10] Hertz (Tong et al., 2017). Therefore, there are mounting evidence that the rsfMRI signal carries perfusion properties, which may be extracted using the novel analytic and modeling approaches.

### 4.2. Better prediction at the parietal lobe or regions

Interestingly, CBF prediction using our approach was better in the parietal lobe and/or regions compared to other areas. The apparent stronger link between rsfMRI activity and perfusion in the parietal area(s) might be related to the performance of our modeling method. However, there could be other reasons as well. We used spectral information of rsfMRI as input features in the model. In fMRI literature, it has been shown that the parietal cortex and/or its sub-regions support a wide range of cognitive processes in association with the prefrontal/frontal cortex. Specifically, studies consistently argue that parietal regions are the key hubs of brain processes and the information is encoded there to communicate with other regions, including prefrontal/frontal regions, and to regulate the cognitive processes (Culham and Valyear, 2006; Fogassi et al., 2005). The parietal lobe is densely vascularized structure (Clark et al., 2017; Duvernoy et al., 1981; Liebeskind, 2003), which might provide a stronger signal to be estimated. In ASL literature, regions within the parietal lobe are consistently reported to be associated with hypoperfusion in MCI or Alzheimer's disease (Alsop et al., 2010; Dolui et al., 2017). The parietal regions thus might be well-positioned for functional measures captured via fMRI and/or ASL perfusion, which eventually could reflect in our machine learning-based predictions.

### 4.3. CBF comparisons between controls and MCI

In our findings, the estimated relative CBF at superior parietal lobule was significantly lower in more impaired MCI patients based on MMSE (< 28) compared to the cognitively normal group. This result was in agreement with prior studies that have reported decreased relative CBF in MCI or Alzheimer's pathology (Dolui et al., 2016; Huang et al., 2018). Interestingly, MCI patients with higher MMSE ($\geq$ 28) had higher estimated perfusion, albeit not statistically significant. This increased perfusion might be related to compensatory mechanism, which has been observed in early stages during the course of Alzheimer's disease (Alsop et al., 2008), and which might be captured more by the rsfMRI.

### 4.4. Implications

Former studies have suggested that ASL captures regional perfusion alterations in patients with Alzheimer's disease (Alsop et al., 2008; Dai et al., 2009). Importantly, ASL has been envisioned as noninvasive and less costly surrogate marker of FDG-PET in the Alzheimer's research (Wolk and Detre, 2012). If ASL data is missing in those populations but rsfMRI is available within with appropriate frequency resolution (i.e., [0.01-0.15] Hertz**)**, it is now possible to impute a surrogate ASL marker at the lobes and/or cortical regions via our new approach. This approach will be useful especially for studies that have large sample size (Habes et al., 2016), where reacquiring missing imaging modalities from the same subject is not practical. CBF estimation was better at the posterior regions, where many previous studies (Alsop et al., 2008; Dolui et al., 2017; Huang et al., 2018) and our results showed altered CBF in MCI or Alzheimer's pathology, suggesting the potential use of proposed method in those populations.

### 4.5. Limitations and future work

Although we were able to estimate CBF at the lobar and cortical ROI levels, especially in the posterior regions, it is worth to note that our method did not offer good perfusion estimates in the subcortical regions. Previous rsfMRI studies have reported that the midbrain and/or nearby subcortical structures are prone to physiological noises such as cardiac cycles (>0.5 Hertz), respiratory cycles (~0.2-0.3 Hertz or lower) (Dagli et al., 1999; Limbrick-Oldfield et al., 2012; Raj et al., 2001). With these cautions, future studies should focus on optimizing perfusion estimates at the subcortical structures.

## 5. Conclusions

In summary, we provided a novel framework to estimate the regional CBF metrics via machine learning algorithm incorporating spectral information of the rsfMRI. This method demonstrated that the perfusion properties can be estimated better using rsfMRI signals with frequencies [0.01-0.15] Hertz compared to signals with frequencies [0.01-0.10] Hertz and [0.01-0.20] Hertz. The perfusion estimation was better in the parietal lobe and/or regions compared to other lobes and/or regions. Furthermore, estimated CBF at the superior parietal lobule significantly associated with cognitive score and it was decreased in MCI with lower cognitive scores. In light of growing attentions of ASL perfusion in aging and Alzheimer's disease pathology, we believe that the proposed approach has the potential to shed light on imputing CBF measures when ASL data are missing.

**Supplementary materials:**

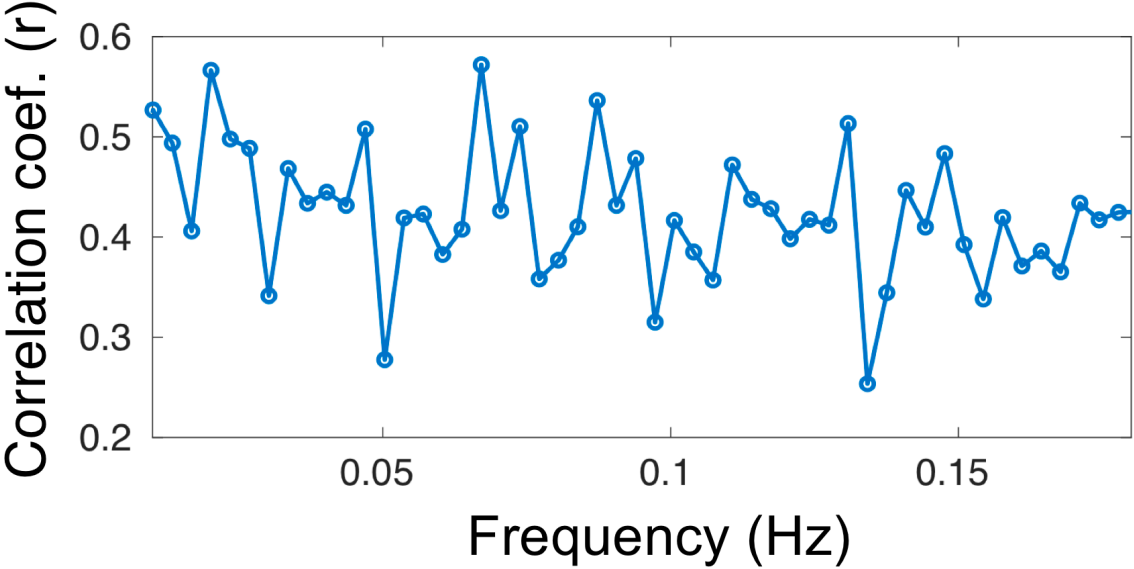

**Figure S1:** Correlation coefficient between predicted CBF and actual CBF at the lobar GM as a function of frequency. Note that each circle represents the correlation coefficient and the circles were connected by a line for visualization.

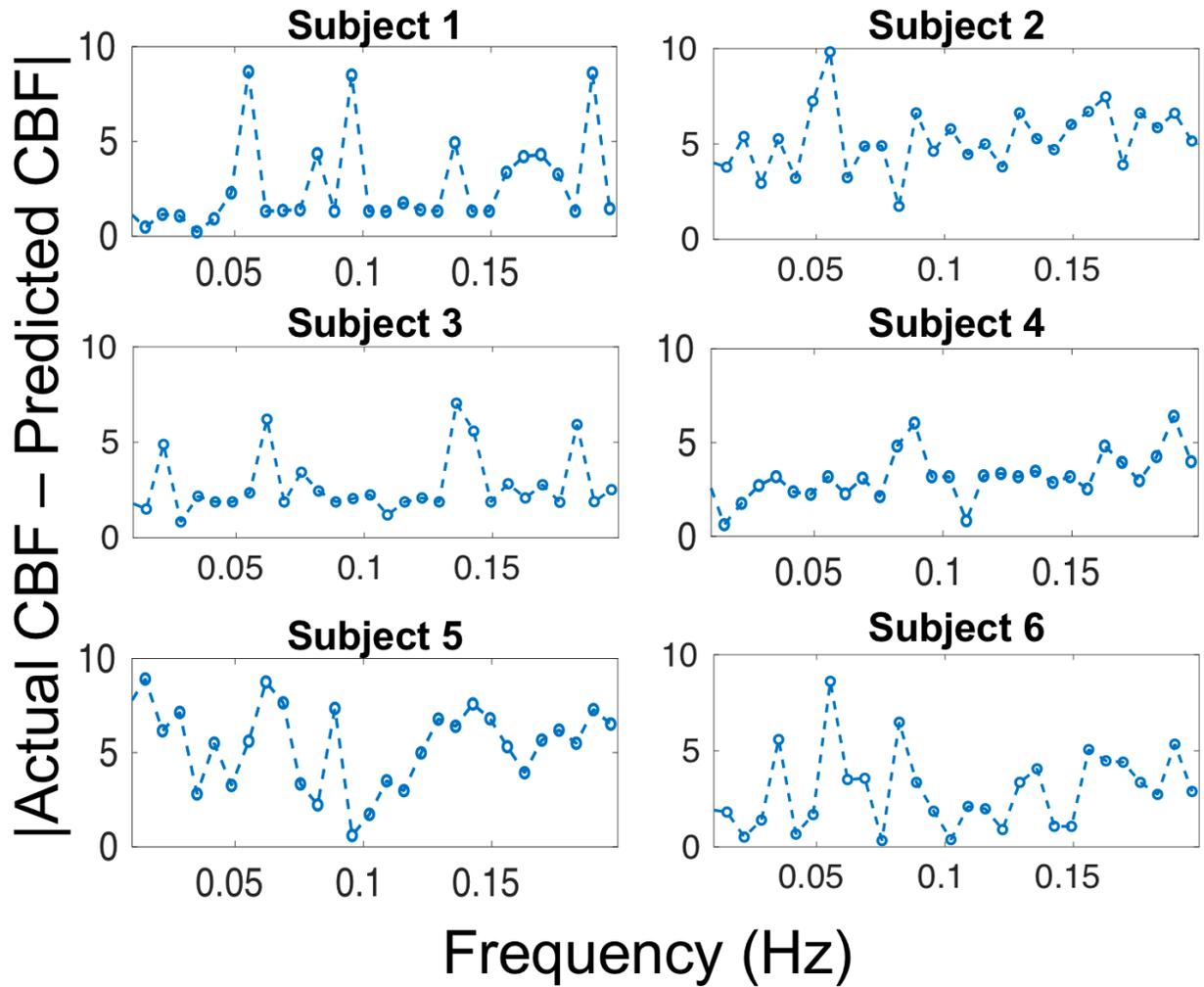

**Figure S2:** Absolute difference between actual CBF and predicted CBF at the lobar GM as a function of frequency in the representative six subjects. Note: Each circle represents the value of absolute difference and the circles were connected for visualization.

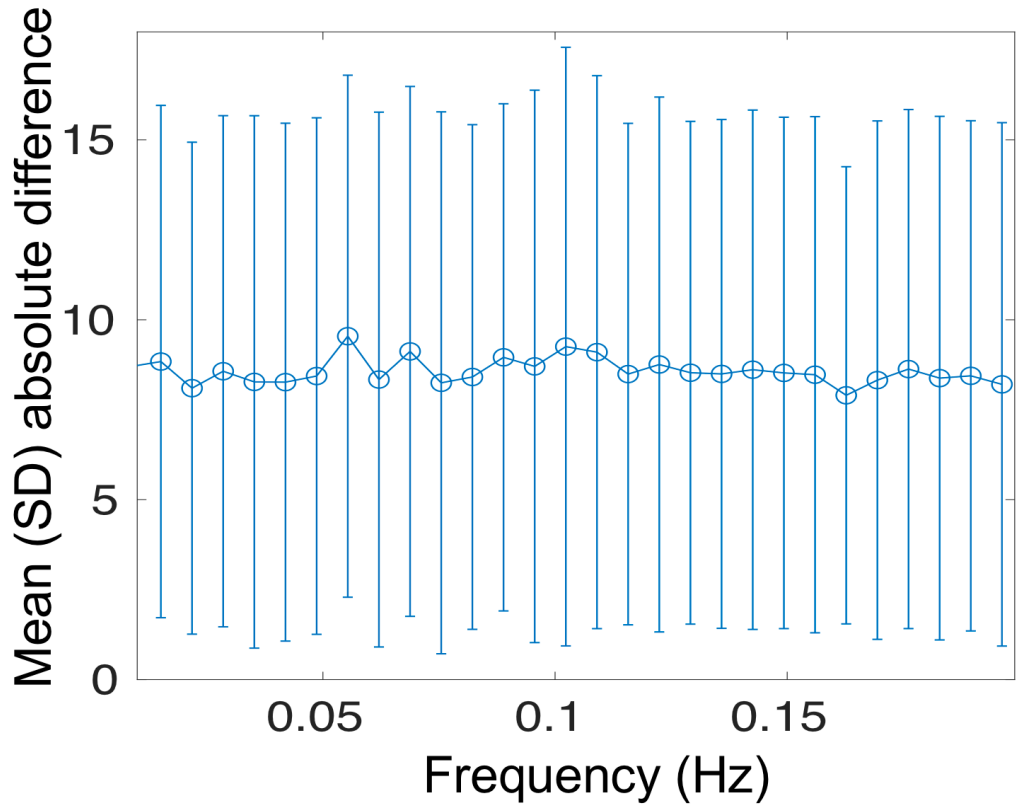

**Figure S3:** Mean [standard deviation (SD)] of absolute difference between actual CBF and predicted CBF over all subjects at the lobar GM as a function of frequency. Note: Each circle represents the mean value of absolute difference and error bar represents SD over all subjects, and the circles were connected for visualization.

**Table S1:** Associations between the actual CBF and predicted CBF at individual ROI level using rsfMRI signals with frequencies [0.01-0.15] Hertz. ROIs with ***bold-italic*** represent they are statistically significant after adjusting for multiple comparisons using FDR.

| ROI name (hemisphere) | Correlation coefficient (r) and p value |
|---|---|
| Angular gyrus (left) | r = 0.34; p = 3.8e-03 |
| ***Cuneus (right)*** | ***r = 0.37; p = 1.6e-03*** |
| ***Inferior occipital gyrus (right)*** | ***r = 0.43; p = 2.0e-04*** |
| ***Inferior occipital gyrus (left)*** | ***r = 0.36; p = 2.2e-03*** |
| Lingual gyrus (left) | r = 0.24; p = 4.8e-02 |
| Middle occipital gyrus (left) | r = 0.26; p = 2.6e-02 |
| Medial orbital gyrus (left) | r = 0.28; p = 1.8e-02 |
| Medial segment of superior frontal gyrus (right) | r = 0.32; p = 5.9e-03 |
| Occipital pole (right) | r = 0.29; p = 1.5e-02 |
| ***Occipital fusiform gyrus (right)*** | ***r = 0.40; p = 6.0e-04*** |
| ***Occipital fusiform gyrus (left)*** | ***r = 0.35; p = 3.2e-03*** |
| Orbital part of inferior frontal gyrus (left) | r = 0.30; p = 9.9e-03 |
| Posterior cingulate gyrus (right) | r = 0.27; p = 2.3e-02 |
| Posterior cingulate gyrus (left) | r = 0.25; p = 3.3e-02 |
| ***Precuneus (right)*** | ***r = 0.35; p = 2.6e-03*** |
| Precuneus (left) | r = 0.31; p = 7.6e-03 |
| Parahippocampal gyrus (left) | r = 0.27; p = 2.1e-02 |
| Subcallosal area (left) | r = 0.29; p = 1.1e-02 |
| Superior frontal gyrus (left) | r = 0.29; p = 1.5e-02 |
| Superior parietal lobule (right) | r = 0.27; p = 2.2e-02 |
| ***Superior parietal lobule (left)*** | ***r = 0.50; p = 9.09e-06*** |
| Caudate (right) | r = 0.24; p = 4.4e-02 |
| Caudate (left) | r = 0.25; p = 3.2e-02 |
| Thalamus proper (left) | r = 0.27; p = 2.3e-02 |